\documentclass{Interspeech}

\title{ParaNoise-SV: Integrated Approach for Noise-Robust Speaker Verification with Parallel Joint Learning of Speech Enhancement and Noise Extraction}

\author[affiliation={1}]{Minu}{Kim}
\author[affiliation={1}]{Kangwook}{Jang}
\author[affiliation={1}]{Hoirin}{Kim}

\affiliation{School of Electrical Engineering}{KAIST}{Republic of Korea}
\email{minus@kaist.ac.kr, dnrrkdwkd12@kaist.ac.kr, hoirkim@kaist.ac.kr}
\keywords{speaker verification, speech enhancement, joint learning, noisy environments, noise disentanglement}

\usepackage[table,xcdraw,dvipsnames]{xcolor}

\usepackage{comment}
\usepackage{multirow} 
\usepackage{booktabs}
\usepackage{makecell}
\usepackage{array}

\begin{document}

\maketitle
\begin{abstract}
Noise-robust speaker verification leverages joint learning of speech enhancement (SE) and speaker verification (SV) to improve robustness. However, prevailing approaches rely on implicit noise suppression, which struggles to separate noise from speaker characteristics as they do not explicitly distinguish noise from speech during training. Although integrating SE and SV helps, it remains limited in handling noise effectively. Meanwhile, recent SE studies suggest that explicitly modeling noise, rather than merely suppressing it, enhances noise resilience. Reflecting this, we propose ParaNoise-SV, with dual U-Nets combining a noise extraction (NE) network and a speech enhancement (SE) network. The NE U-Net explicitly models noise, while the SE U-Net refines speech with guidance from NE through parallel connections, preserving speaker-relevant features. Experimental results show that ParaNoise-SV achieves a relatively 8.4\% lower equal error rate (EER) than previous joint SE-SV models.
\end{abstract}

\section{Introduction}

Rapid advancement of modern communication devices and voice-based technologies has highlighted the growing importance of speaker verification\,(SV) systems.
These systems, whi-ch verify whether a given speech matches a target speaker, are critical for applications such as secure authentication and forensic analysis\,\cite{ singh2012applications}.
However, real-world environments are rarely free of noise, posing a major challenge for speaker verification systems\,\cite{song2018noise}.
Conventional solutions integrate separately trained speech enhancement (SE) modules to mitigate noise\,\cite{plchot2016audio, kolboek2016speech}, however, they can degrade speaker-specific information, leading to suboptimal embeddings and reduced verification accuracy \cite{ma2021pl}.
Furthermore, independently trained SE and SV modules lack coherence with each other, as SE outputs may not align well with the learned feature distributions of the SV system \cite{wu2021joint}.

To address these challenges, several studies have explored joint learning strategies for SE and SV.
A previous study \cite{shi2020robust} integrates SE with an attention-based model, while another approach \cite{wu2021joint} introduces a multi-objective network for simultaneous feature enhancement and speaker embedding extraction.
A U-Net-based approach \cite{kim2022extended} jointly optimizes SE and SV, and diffusion models \cite{kim2024diff} disentangle speaker and noise representations for improved robustness.
Likewise, a noise-disentanglement adversarial framework \cite{xing2024joint} separates speaker-relevant and irrelevant information for robust embeddings in noise.
Furthermore, a recent study \cite{lim2024noise} incorporates noise type and signal-to-noise ratio\,(SNR) level into SE, demonstrating the benefits of noise estimation.
Meanwhile, self-supervised learning\,(SSL)-based verification \cite{lim2024improving} offers an alternative approach to improving noise robustness by fine-tuning large-scale pre-trained models.
Despite its effective performance, its larger parameter size increases latency and computational demands.

Beyond its role in joint learning with SV, SE has also been extensively studied as a standalone approach to improve speech robustness. Rather than relying solely on predefined noise attributes (e.g., amplitude and SNR) to guide enhancement \cite{lee2020dynamic, deng2020naagn}, recent SE methods focus on dynamically synthesizing noise representations within the SE network itself. For instance, a neural noise embedding approach \cite{zhang2021neural} generates noise representations and applies conditional encoding with layer normalization, improving speech quality in end-to-end systems. Similarly, dual-stream SE models \cite{lu2023speech, li2025dual} process noise and speech separately, resulting in better enhancement.

\begin{figure}[!t]
  \centering
  \includegraphics[width=\linewidth]{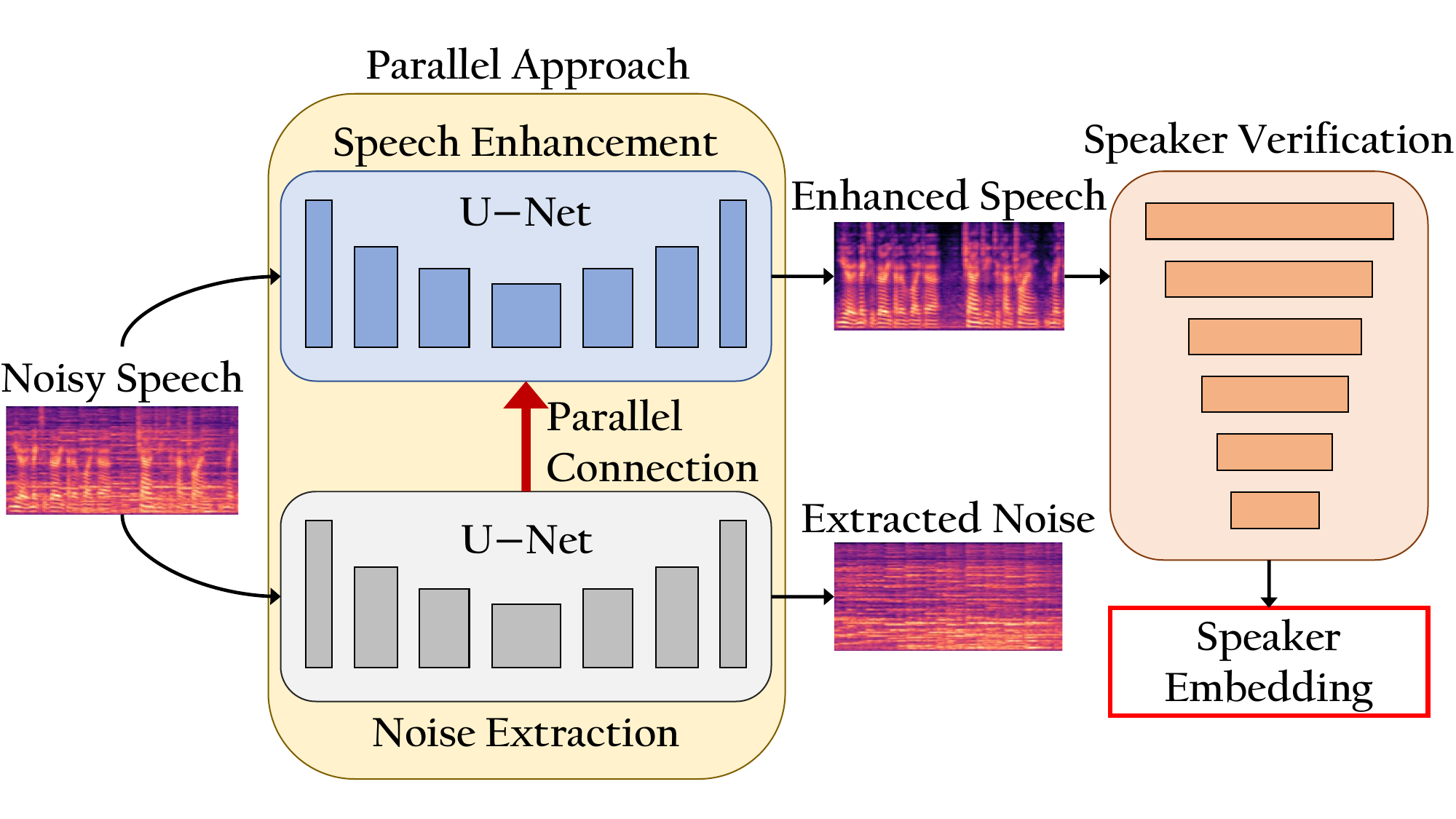}
  \vspace{-15pt}
  \caption{Overview of ParaNoise-SV architecture. Noise extraction\,(NE) network and speech enhancement (SE) network are trained in parallel and connected via parallel connections. The enhanced speech is then processed by speaker verification\,(SV) module for robust speaker embedding extraction.}
  \label{fig:model1}
  \vspace{-10pt}
\end{figure}

Building on these approaches, we propose \textbf{ParaNoise-SV}, a unified framework for noise-robust speaker verification. Instead of merely predicting coarse noise attributes, our model dynamically extracts and synthesizes background noise from speech to effectively remove speaker-irrelevant components. To achieve this, we design a noise extraction (NE) network and a speech enhancement (SE) network, both based on U-Net \cite{ronneberger2015u}, and train them simultaneously. These two networks are interconnected through parallel connections, allowing noise-related information to refine speech while preserving speaker-relevant features, as in Figure \ref{fig:model1}. By jointly learning NE, SE, and speaker verification (SV), our system improves robustness against noise while maintaining high speaker discrimination. ParaNoise-SV achieves relatively 8.4\% lower EER than previous joint SE-SV models in seen noise conditions and reduces EER by 8.2\% in unseen noise conditions.

\section{Methods}
\subsection{Overview of ParaNoise-SV}
\begin{figure}[!t]
  \centering
  \includegraphics[width=\linewidth]{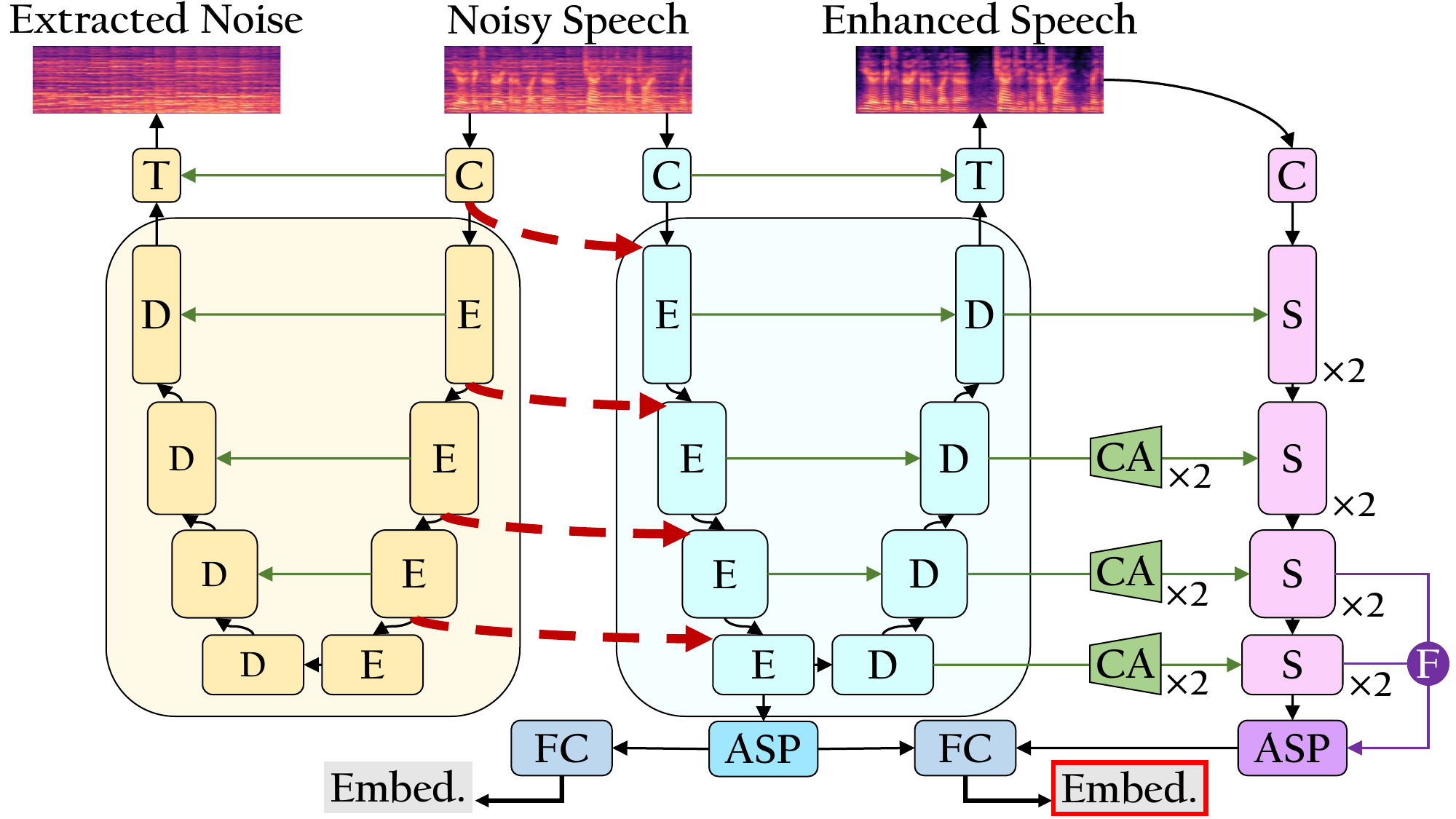}
  \vspace{-15pt}
  \caption{Dual U-Nets for SE and NE networks are trained in parallel with parallel connections (dashed red line) for noise suppression. They use E and D layers based on SE-ResNet. The SV network processes enhanced speech using S layers from ERes2NetV2, with CA for channel alignment in skip connections and F for multi-channel fusion. Speaker embeddings are extracted in two stages via attentive statistics pooling\,(ASP). (C: convolution, T: transposed convolution, E: encoding, D: decoding, S: speaker extraction, F: bottom-up fusion, CA: channel adaptation)
  }
  \label{fig:model2}
\end{figure}

\begin{figure}[!t]
  \centering
  \includegraphics[width=\linewidth]{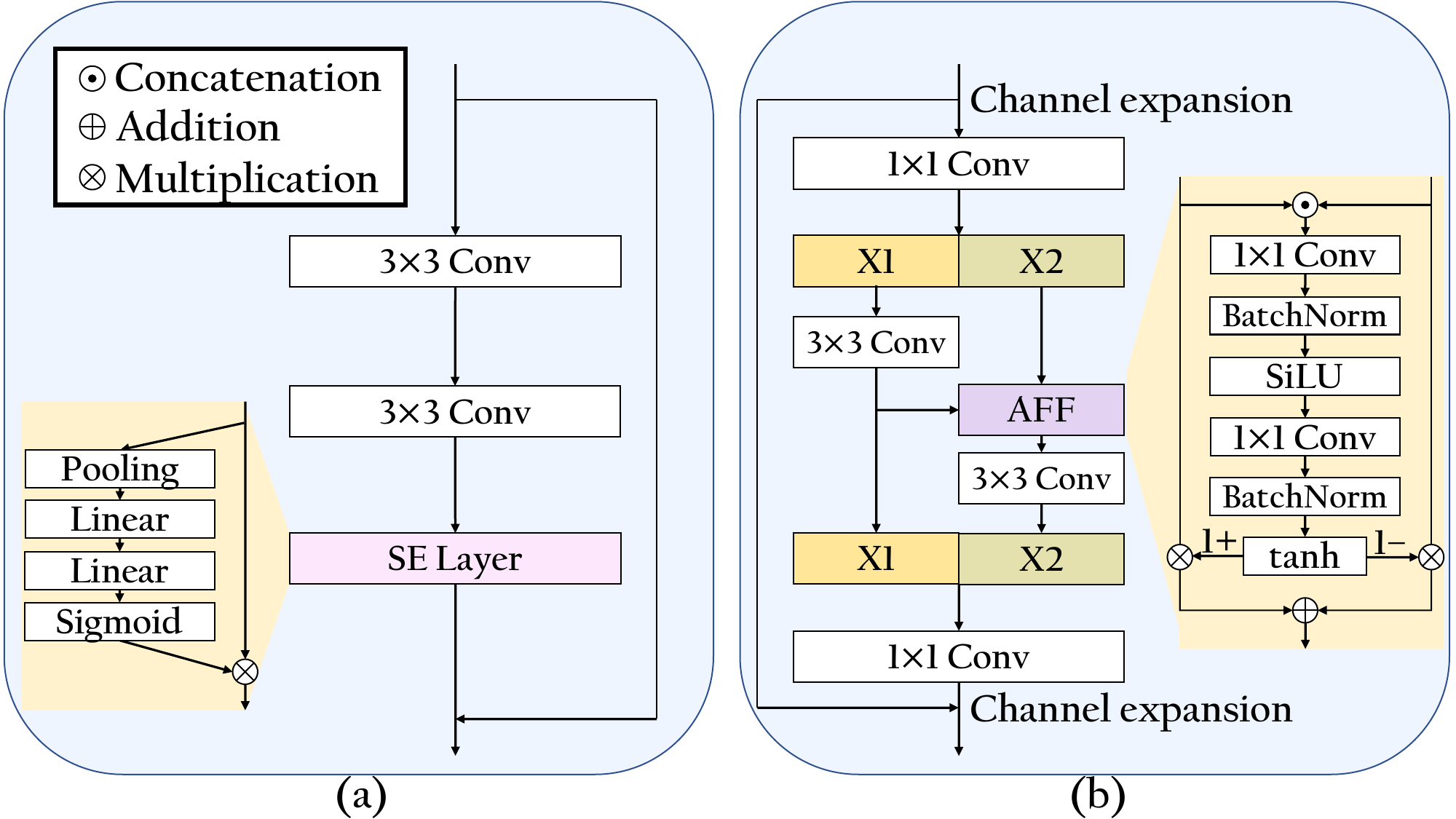}
  \vspace{-15pt}
  \caption{(a) SE-ResNet applies squeeze-and-excitation mechanisms to enhance important feature channels by adaptively reweighting them. (b) ERes2NetV2 incorporates channel-split hierarchical residual connections with expanded channel dimensions, improving multi-scale feature extraction.}
  \label{fig:model3}
  \vspace{-10pt}
\end{figure}

ParaNoise-SV is a unified framework for noise-robust speaker verification, integrating NE, SE, and SV using dual encoder-decoder structures.
It employs dual U-Nets with SE-ResNet \cite{hu2018squeeze} for simultaneous NE and SE, introducing parallel connections to ensure balanced separation while preserving speaker-relevant information.  
The NE network isolates noise, which the SE network leverages through parallel connections at each encoding stage, enabling dynamic noise suppression while maintaining speaker discriminability. Unlike conventional methods focusing solely on suppression or enhancement, ParaNoise-SV actively utilizes extracted noise at the feature level, preventing contamination in deeper representations.  
For speaker embedding extraction, ERes2NetV2 \cite{chen2024eres2netv2} is used, with channel adaptation blocks integrating the U-Net features via skip connections. The key frameworks are shown in Figures \ref{fig:model2} and \ref{fig:model3}.

\subsection{Parallel Connections of Dual U-Nets}

The input spectrogram is first normalized using instance normalization \cite{ulyanov2016instance} and processed through an initial convolutional layer, generating noise and speech feature maps \( N_{E,0} \) and \( S_{E,0} \). Each encoder then extracts hierarchical representations using SE-ResNets \cite{hu2018squeeze} with depth \( L=4 \).  

\vspace{-10pt}
\begin{align}
    N_{E,i} &= e_{N}^i(N_{E,i-1}), \quad i = 1, \dots, L.
\label{eq:squeeze_noise}\\
    N_{D,0} &= N_{E,L}.
\label{eq:extend0_noise}\\
    N_{D,i} &= d_N^i(N_{D,i-1},N_{E,L-i}), \quad i = 1, \dots, L.
\label{eq:extend_noise}\\
    \hat{N} &= \text{ConvTranspose}(N_{D,L},N_{E,0}).
\label{eq:enhanced_noise}
\end{align}

\noindent The NE network encodes noise representations through encoder blocks \( e_N \) in Equation \eqref{eq:squeeze_noise}, refining noise features. The deepest encoded feature in Equation \eqref{eq:extend0_noise} initializes the decoding operations \( d_N \), where skip connections aid noise extraction as in Equation \eqref{eq:extend_noise}. Finally, a transposed convolutional layer generates the estimated noise spectrogram \( \hat{N} \) in Equation \eqref{eq:enhanced_noise}.

\vspace{-10pt}
\begin{align}
    S_{E,i} &= e_{S}^i(S_{E,i-1},N_{E,i-1}), \quad i = 1, \dots, L
\label{eq:squeeze_speech}\\
    S_{D,0} &= S_{E,L}
\label{eq:extend0_speech}\\
    S_{D,i} &= d_S^i(S_{D,i-1},S_{E,L-i}), \quad i = 1, \dots, L
\label{eq:extend_speech}\\
    \hat{S} &= \text{ConvTranspose}(S_{D,L},S_{E,0})
\label{eq:enhanced_speech}
\end{align}

\noindent In Equation \eqref{eq:squeeze_speech}, the SE network incorporates \textit{parallel connections}: information flows between two parallel networks, integrating noise features at each encoder block \( e_S \). The encoded speech feature is decoded with skip connections as in Equation \eqref{eq:extend_speech}, and a final transposed convolutional layer outputs the enhanced speech spectrogram \( \hat{S} \) in Equation \eqref{eq:enhanced_speech}. This allows the SE network to utilize noise information from the NE network, improving noise suppression while preserving speaker details.

\subsection{Speaker Embedding Extraction}

In noisy speech, spectral components can become missing or corrupted, making speaker verification challenging. Capturing information at multiple scales helps preserve speaker characteristics even when specific frequency bands are degraded.
To address this, we utilize ERes2NetV2 \cite{chen2024eres2netv2}, originally designed for short-duration speaker verification, where extracting features from limited speech is critical. It tackles the challenge of incomplete temporal or spectral information by employing multi-scale feature fusion and channel expansion, effectively capturing speaker-relevant details across varying time and frequency resolutions.
Leveraging these strengths, ERes2NetV2 enhances speaker verification in noisy environments by recovering speaker characteristics from degraded spectral components.

To fully exploit the speech refinement capability of SE network, we integrate its decoder features into ERes2NetV2 via skip connections, allowing ERes2NetV2 to adjust channel importance to enhance speaker-relevant information.
However, ERes2NetV2 expands channels by 2, as in Figure~\ref{fig:model2}, causing a dimension mismatch that hinders direct integration. To resolve this, channel adaptation, consisting of multiple convolution layers, is applied to SE decoder outputs.

Speaker embedding extraction follows a two-stage process to minimize information loss caused by noise suppression. First, the deepest encoder feature of SE network undergoes ASP \cite{okabe2018attentive}, becoming the initial pooled vector, which then passes through a fully connected (FC) layer to generate an initial embedding. After further processing through the SV network, ASP is applied again, and the refined features are concatenated with the initial pooled vector before passing through a final FC layer. This final embedding utilizes the output of the SE encoder, refined through parallel connections, with the output from ERes2NetV2, achieved through multi-scale aggregation. This integration enhances noise robustness and preserves detailed and global speaker characteristics, providing comprehensive representation and improved discrimination performance. The final embedding\,(red-edged box in Figure\,\ref{fig:model2}) is ultimately used as the identity representation for speaker verification.

\subsection{Loss Functions}

\begin{equation}
    L = L_{n} + L_{s} + L_{C} + L_{\text{AP}} + L_{\text{AAM}}
    \label{loss}
\end{equation}

\noindent The loss function in Equation~\eqref{loss} optimizes NE, SE, and SV in an integrated framework. The noise extraction loss \( L_n \) is formulated as mean squared error (MSE) loss between \( \hat{N} \) and the original noise spectrogram. This promotes precise noise extraction and minimizes residual noise interference in speaker embeddings, reducing leakage into enhanced speech.
The speech enhancement loss \( L_s \) also applies MSE loss between \( \hat{S} \) and the clean speech spectrogram, preserving speech intelligibility.

For speaker embedding extraction, cross-entropy loss (\( L_{\text{C}} \)) is applied between the initial embedding and its speaker label. The final embedding is optimized using angular prototypical (AP) loss (\( L_{\text{AP}} \))~\cite{zhou2019deep} and additive angular margin (AAM) loss (\( L_{\text{AAM}} \))~\cite{deng2019arcface}. To enhance noise robustness, each batch contains both clean and noisy speech from randomly selected speakers, as proposed in~\cite{kim2022extended}. \( L_{\text{AP}} \) aligns final clean and noisy embeddings of the same speaker, while \( L_{\text{AAM}} \) separates final speaker embeddings from their speaker-wise prototypes.

\begin{table}[t!]
    \caption{Architectures of the SE and SV networks.
The SE network consists of an encoder (E) and a decoder (D), with each layer (L) in the encoder applying concatenation before SE-ResNet units for parallel connections, while the decoder follows a residual structure.
The NE network shares the SE structure but omits parallel connections, excluding concatenation and convolution in the encoder.
The SV network uses Res2Net and ERes2NetV2 blocks for multi-scale feature extraction, doubling output channels via expansion parameters.
TConv2d denotes transposed convolution layers, with output channel dimensions (Ch.) shown at the bottom of each block.}
    \label{tab:structures}
\resizebox{\linewidth}{!}{%
    \begin{tabular}{c|c||c|c||c|c}
    \Xhline{2\arrayrulewidth}
    \multirow{2}{*}{\textbf{L}} & \textbf{SE Encoder} & \multirow{2}{*}{\textbf{L}} & \textbf{SE Decoder}& \multirow{2}{*}{\textbf{L}} & \textbf{SV} \\
     & \textbf{Structure} & & \textbf{Structure} & & \textbf{Structure} \\
    \Xhline{2\arrayrulewidth}
    C  &
        \begin{tabular}{c}
            Conv2d\\
            \hline
            Ch. = 16
        \end{tabular} 
    & T  &
        \begin{tabular}{c}
            Concat \\
            TConv2d \\
            \hline
            Ch. = 1
        \end{tabular} 
    & C  &
        \begin{tabular}{c}
            Conv2d \\
            \hline
            Ch. = 16
        \end{tabular} \\
    \Xhline{\arrayrulewidth}
    E1 &
        \begin{tabular}{c}
            Concat \\
            Conv2d \\
            SE-ResNet ×3 \\
            \hline
            Ch. = 16
        \end{tabular} 
    & D4 &
        \begin{tabular}{c}
            Concat \\
            Conv2d \\
            SE-ResNet ×3 \\
            \hline
            Ch. = 16
        \end{tabular} 
    & S1 &
        \begin{tabular}{c}
            Concat \\
            Conv2d \\
            Res2Net ×3 \\
            \hline
            Ch. = 16 → 32
        \end{tabular} \\
    \Xhline{\arrayrulewidth}
    E2 &
        \begin{tabular}{c}
            Concat \\
            Conv2d \\
            SE-ResNet ×4 \\
            \hline
            Ch. = 32
        \end{tabular} 
    & D3 &
        \begin{tabular}{c}
            Concat \\
            TConv2d \\
            SE-ResNet ×4 \\
            \hline
            Ch. = 16
        \end{tabular} 
    & S2 &
        \begin{tabular}{c}
            Concat \\
            Conv2d \\
            Res2Net ×4 \\
            \hline
            Ch. = 32 → 64
        \end{tabular} \\
    \Xhline{\arrayrulewidth}
    E3 &
        \begin{tabular}{c}
            Concat \\
            Conv2d \\
            SE-ResNet ×6 \\
            \hline
            Ch. = 64
        \end{tabular} 
    & D2 &
        \begin{tabular}{c}
            Concat \\
            TConv2d \\
            SE-ResNet ×6 \\
            \hline
            Ch. = 32
        \end{tabular} 
    & S3 &
        \begin{tabular}{c}
            Concat \\
            Conv2d \\
            ERes2NetV2 ×6 \\
            \hline
            Ch. = 64 → 128
        \end{tabular} \\
    \Xhline{\arrayrulewidth}
    E4 &
        \begin{tabular}{c}
            Concat \\
            Conv2d \\
            SE-ResNet ×3 \\
            \hline
            Ch. = 128
        \end{tabular} 
    & D1 &
        \begin{tabular}{c}
            Concat \\
            Conv2d \\
            SE-ResNet ×3 \\
            \hline
            Ch. = 64
        \end{tabular} 
    & S4 &
        \begin{tabular}{c}
            Concat \\
            Conv2d \\
            ERes2NetV2 ×3 \\
            \hline
            Ch. = 128 → 256
        \end{tabular} \\
    \Xhline{2\arrayrulewidth}
    \end{tabular}
}
\vspace{-10pt}
\end{table}

\section{Experimental Setup}

\begin{table*}[t!]
\caption{
Experimental results (EER \%) on the VoxCeleb1 test set with noise scenarios synthesized from the MUSAN corpus at various SNR levels. For our model, we conduct a performance evaluation under four conditions: without parallel connections, with the connections to the decoder only, with the connections to both the encoder and decoder, and with the connections to the encoder only.
}
\vspace{-5pt}
\centering
\label{table:result1}
\resizebox{\textwidth}{!}{
\begin{tabular}{c|c|c|c|c|c|c|c|c|c|c|c|c|c|c|c|c|c|c} \hline\hline
\multirow{3}{*}{Method} & \multirow{3}{*}{Models} & \multicolumn{17}{c}{EER (\%)} \\ \cline{3-19}
 &  & \multirow{2}{*}{Clean} & \multicolumn{5}{c|}{Babble} & \multicolumn{5}{c|}{Music} & \multicolumn{5}{c|}{Noise} & \multirow{2}{*}{Avg.} \\\cline{4-18}
 &  & & 0 & 5 & 10 & 15 & 20 & 0 & 5 & 10 & 15 & 20 & 0 & 5 & 10 & 15 & 20 & \\ \hline
\multirow{8}{*}{\shortstack{Joint\\SE + SV}}
& VoiceID~\cite{shon2019voiceid} & 6.79 & 37.96 & 27.12 & 16.66 & 11.25 & 8.99 & 16.24 & 11.44 & 9.13 & 8.10 & 7.48 & 16.56 & 12.26 & 9.86 & 8.69 & 7.83 & 13.52 \\
& Shi \textit{et al.}~\cite{shi2020robust} & 6.18 & 37.55 & 26.42 & 16.30 & 10.89 & 8.39 & 15.58 & 10.93 & 8.87 & 7.62 & 7.13 & 15.95 & 11.76 & 9.17 & 8.08 & 7.07 & 12.99 \\
& Wu \textit{et al.}~\cite{wu2021joint}     & 7.60                   & 20.11 & 12.02 & 9.63  & 8.48 & 7.99 & 12.92 & 10.10 & 8.95 & 8.35 & 7.95 & 13.12 & 10.57 & 9.28 & 8.59 & 8.10 & 10.24 \\
& NDML~\cite{sun2023noise}  & 2.90                   & 10.96 & 6.13  & 4.28  & 3.52  & 3.21 & 10.84 & 6.52  & 4.66 & 3.67 & 3.21 & 10.24 & 6.96  & 5.02 & 3.91 & 3.40 & 5.59 \\
 & ExU-Net~\cite{kim2022extended} & 2.76                   & 9.57  & 5.52  & 4.06  & 3.28 & 2.99 & 7.35  & 4.90  & 3.69 & 3.14 & 2.93 & 6.80  & 5.23  & 4.07 & 3.39 & 3.10 & 4.55 \\
& Diff-SV~\cite{kim2024diff} & 2.35 & 8.74 & 4.51 & 3.33 & 2.82 & 2.61 & 6.04 & 3.96 & 3.10 & 2.75 & 2.60 & 6.01 & 4.52 & 3.49 & 2.93 & 2.64 & 3.90 \\
& NDAL~\cite{xing2024joint}  & 2.63                   & 6.14 & 4.00  & 3.23  & 2.97  & 2.80 & 6.43 & 4.44  & 3.59 & 3.08 & 2.87 & 5.87 & 4.19  & 3.53 & 3.23 & 3.09 & 3.88 \\
 & NA-ExU-Net~\cite{lim2024noise} & 1.99 & 9.88  & 4.57  & 3.10  & 2.43 & 2.09 & 6.24  & 3.95  & 2.80 & 2.39 & 2.17 & 5.85 & 3.90  & 3.05 & 2.54 & 2.37 & 3.71 \\
\hline
\multirow{4}{*}{\textbf{\textit{Ours}}}
& \textbf{Baseline (\textit{w/o NE})}  & 2.23 & 10.03 & 4.85 & 3.01 & 2.72 & 2.35 & 7.01 & 3.98 & 2.94 & 2.64 & 2.32 & 6.01 & 4.05 & 3.27 & 2.59 & 2.46 & 3.90 \\
& \textbf{ParaNoise-SV (\textit{dec.})} & 2.03 & 10.24 & 4.73 & 3.13 & 2.78 & 2.43 & 7.06 & 4.22 & 3.24 & 2.68 & 2.35 & 6.12 & 4.08 & 3.36 & 2.72 & 2.41 & 3.97 \\

& \textbf{ParaNoise-SV (\textit{enc., dec.})}  & 1.87 & 9.46 & 4.40 & 2.75 & 2.21 & 1.94 & 6.60 & 3.71 & 2.58 & 2.03 & 1.92 & 5.85 & 3.74 & 2.76 & 2.25 & 2.02 & 3.51 \\
& \cellcolor[HTML]{f9caf1}\textbf{ParaNoise-SV (\textit{enc.})}  & \cellcolor[HTML]{f9caf1}\textbf{1.75}  & \cellcolor[HTML]{f9caf1}9.46 & \cellcolor[HTML]{f9caf1}4.37  & \cellcolor[HTML]{f9caf1}\textbf{2.64}  & \cellcolor[HTML]{f9caf1}\textbf{2.13}  & \cellcolor[HTML]{f9caf1}\textbf{1.84} & \cellcolor[HTML]{f9caf1}6.20 & \cellcolor[HTML]{f9caf1}\textbf{3.62}  & \cellcolor[HTML]{f9caf1}\textbf{2.46} & \cellcolor[HTML]{f9caf1}\textbf{1.95} & \cellcolor[HTML]{f9caf1}\textbf{1.81} & \cellcolor[HTML]{f9caf1}\textbf{5.60} & \cellcolor[HTML]{f9caf1}\textbf{3.64}  & \cellcolor[HTML]{f9caf1}\textbf{2.75} & \cellcolor[HTML]{f9caf1}\textbf{2.12} & \cellcolor[HTML]{f9caf1}\textbf{2.00} & \cellcolor[HTML]{f9caf1}\textbf{3.40} \\\hline\hline
\end{tabular}
}
\vspace{-10pt}
\end{table*}

\begin{table}[t!]
\caption{
   Experimental results (EER \%) on the VoxCeleb1 test set with an out-of-domain noise source (NonSpeech100).
}
\vspace{-5pt}
\centering
\label{table:out}
\resizebox{\linewidth}{!}{
\begin{tabular}{c|c c c c c c} \hline\hline
SNR                                           & NDML  & ExU-Net  & Diff-SV   & NDAL      & NA-ExU-Net       & \textbf{ParaNoise-SV} \\ \hline
0                             & 20.49 & 8.39   & 8.23  & 7.57  & 7.82          & 7.61 \\
5                             & 15.09 & 5.59   & 5.06  & 5.49  & 4.78          & \textbf{4.40} \\
10                            & 11.96 & 4.36   & 3.85  & 4.03  & 3.46          & \textbf{3.17} \\
15                            & 9.96  & 3.74   & 3.19  & 3.36  & 2.75          & \textbf{2.22} \\
20                            & 8.64  & 3.29   & 2.89  & 2.99  & 2.44          & \textbf{2.09} \\\hline
Avg.                          & 13.23 & 5.01   & 4.64  & 4.69  & 4.25          & \textbf{3.90} \\\hline \hline
\end{tabular}
 }
 \vspace{-5pt}
\end{table}

\begin{table}[t!]
\caption{
   Average EER (\%) on the VoxCeleb1 test set.  
   The seen condition includes clean speech and noisy speech with MUSAN, while the unseen condition uses noisy speech with NonSpeech100.
}
\vspace{-5pt}
\centering
\label{table:out2}
\resizebox{\linewidth}{!}{
\begin{tabular}{c|>{\centering\arraybackslash}m{1.5cm} >{\centering\arraybackslash}m{1.5cm} >{\centering\arraybackslash}m{1.5cm}|c} 
\hline\hline
\multirow{2}{*}{Noise Type} & \multicolumn{3}{c|}{Counterpart (noise attribute estimation)} & \multirow{2}{*}{\textbf{ParaNoise-SV}} \\ \cline{2-4}
             & Class & SNR & Class+SNR &  \\ \hline
Seen         &  3.80  & 3.69  & 3.66  & \textbf{3.40} \\
Unseen       &  4.47  & 4.34  & 4.31  & \textbf{3.90} \\\hline \hline
\end{tabular}
 }
\vspace{-10pt}
\end{table}

\textbf{Model Structure Details.} As shown in Table~\ref{tab:structures}, the SE network employs parallel connections, where noise features from the NE network are concatenated at each encoder stage before downsampling. This is reflected in the encoder layers (E1–E4), where concatenation occurs before convolutional processing, enabling adaptive noise suppression while preserving speaker-relevant features. In contrast, the NE network omits these connections to maintain independent noise modeling. The decoder (D1–D4) mirrors the encoder, incorporating residual connections to reconstruct the enhanced speech effectively.

The SV network integrates Res2Net blocks (S1, S2) \cite{gao2019res2net} for fine-grained spectral and temporal feature extraction and ERes2NetV2 blocks (S3, S4) \cite{chen2024eres2netv2} for deeper multi-scale integration. To enhance feature learning across scales, adaptive feature fusion (AFF) in Figure~\ref{fig:model3}(b) is applied only to the third and fourth layers. This ensures that early layers (S1, S2) focus on capturing local speech details, while deeper layers (S3, S4) integrate broader speaker identity features.

To further stabilize speaker representations, the final embedding is obtained through a bottom-up fusion \cite{chen2024eres2netv2} as in Figure~\ref{fig:model2}, which follows a process similar to AFF. In this process, the third-layer output is first restructured before being combined with the fourth-layer output. This combination allows intermediate temporal details and high-level speaker identity features to be effectively integrated. The fusion enhances noise resilience and leads to a more stable speaker representation.
 \vspace{3pt} \\
\textbf{Training Details.} The model is trained for 200 epochs with the Adam optimizer with a weight decay of 1e-4 and AAM loss with a margin of 0.15 and a scale of 32. The learning rate peaks at 0.01 over 5 warm-up epochs and decays via cosine annealing. Input features are 64-dimensional log Mel-spectrograms (25 ms window, 10 ms hop), with SpecAugment~\cite{park2019specaugment}. The final speaker embedding dimension is 192. Performance is measured using equal error rate (EER) on clean and noisy datasets.
\vspace{3pt} \\
\textbf{Datasets.} ParaNoise-SV is evaluated on the VoxCeleb1 dataset~\cite{nagrani2017voxceleb}, with 148,642 utterances from 1,211 speakers for training and 4,874 utterances from 40 speakers for testing.
Models are trained with noisy speech by mixing MUSAN~\cite{snyder2015musan} noise at randomly chosen SNR between 0 and 20 dB, with speed perturbation applied to the raw waveforms.
 Testing is conducted at $\{0, 5, 10, 15, 20\}$ dB.
Additional robustness is assessed using NonSpeech100~\cite{hu2010tandem} noise dataset.

\section{Results}

\subsection{Main Results}
Table~\ref{table:result1} presents the EER results for ParaNoise-SV under clean and noisy conditions at different SNR levels. The results dem-onstrate that ParaNoise-SV consistently achieves lower EERs across all noise conditions, indicating strong noise robustness. In clean conditions, it achieves an EER of 1.75\%, confirming its effectiveness in preserving speaker identity. When averaging across both clean and noisy speech, ParaNoise-SV achieves an overall EER of 3.40\%, outperforming previous joint SE-SV approaches for noise-robust speaker verification. This suggests that even under demanding noisy conditions, ParaNoise-SV effectively balances noise disentanglement and speaker identity preservation, addressing the limitations of conventional joint learning frameworks that rely on implicit noise suppression.  

Table~\ref{table:out} presents the VoxCeleb1 test results under an out-of-domain noise scenario using NonSpeech100. Even in this challenging setting, ParaNoise-SV achieves the lowest average EER of 3.90\%, demonstrating better generalization compared to existing joint learning models. This further supports its effectiveness in explicit noise extraction, ensuring robust performance across both seen and unseen noise conditions.  
\subsection{Discussion}  
\textbf{Effect of Parallel Connections.} To analyze the impact of parallel connections, we conduct an ablation study comparing different parallel connection setups. The baseline model follows a conventional joint learning setup with a single U-Net and no explicit noise extraction, maintaining the overall structure of ParaNoise-SV but without the NE network or parallel connections. We evaluate three variants: one with decoder-to-decoder parallel connections only (ParaNoise-SV (dec.)), another with both encoder-to-encoder and decoder-to-decoder connections (ParaNoise-SV (enc., dec.)), and a third with encoder-to-encoder connections only (ParaNoise-SV (enc.)).
The results confirm that encoder-level parallel connections significantly enhance noise disentanglement, improving speaker verification robustness. ParaNoise-SV (enc.) achieves the lowest average EER of 3.40\%, demonstrating that propagating noise information at the encoder stage is the most effective strategy for preserving speaker-relevant features.

On the other hand, the results also reveal the impact of different parallel connection strategies. Applying parallel connections only at the decoder (ParaNoise-SV (dec.)) leads to performance degradation over the baseline. This indicates that introducing noise-related information at a later stage disrupts feature refinement by interfering with the learned representations of the enhanced speech. Even when encoder-level connections are present, adding decoder-level connections (ParaNoise-SV (enc., dec.)) further degrades performance compared to encoder-only connections. This reinforces that incorporating noise-related information at later stages hinders disentanglement rather than enhancing it.
\vspace{3pt} \\
\textbf{Comparison with Noise Attribute Estimation.} To better illustrate the significance of noise synthesis in parallel noise extraction models, we compare ParaNoise-SV with variants that estimate noise attributes instead of synthesizing noise. These counterparts retain the same architecture but replace the NE network’s decoder with a noise attribute prediction layer, which infers noise class and SNR as \cite{lim2024noise}. As Table~\ref{table:out2} shows, ParaNoise-SV outperforms the variants in both seen and unseen environments, highlighting the benefits of explicit noise modeling.
\vspace{3pt} \\
\textbf{Comparison with SSL.} We next compare ParaNoise-SV with SSL-based speaker verification models \cite{lim2024improving}, which leverage large-scale pre-trained models and noise-adaptive fine-tuning. As shown in Table~\ref{table:param_eer}, despite having significantly fewer parameters, ParaNoise-SV outperforms HuBERT + NAW-SV in both seen and unseen conditions.
While HuBERT + NAW-SV has a large parameter count, it struggles in noisy conditions, as pre-training objectives of HuBERT \cite{hsu2021hubert} are less aligned with noise robustness.
In contrast, WavLM + NAW-SV benefits from the strengths of WavLM \cite{chen2022wavlm} itself, including speech denoising pre-training and diverse augmentations. As a result, the performance gap between HuBERT + NAW-SV and WavLM + NAW-SV highlights the dependency of SSL-based models on their respective foundation models. Nonetheless, ParaNoise-SV achieves competitive performance with a model size over 13 times smaller, demonstrating the model's efficiency towards noise-robustness without relying on large-scale pre-training.

\begin{table}[t!]
\caption{
   Model size and average EER (\%) on the VoxCeleb1 test set, comparing with SSL-based verification. The seen condition includes clean speech and noisy speech with MUSAN, while the unseen condition uses NonSpeech100.
}
\vspace{-5pt}
\centering
\label{table:param_eer}
\resizebox{\linewidth}{!}{
\begin{tabular}{l|>{\centering\arraybackslash}m{2.2cm}| >{\centering\arraybackslash}m{1.2cm} >{\centering\arraybackslash}m{1.2cm}} 
\hline\hline
Model                  & \# Parameters  & Seen & Unseen \\ \hline
HuBERT + NAW-SV \cite{lim2024improving}   & 102M+          & 4.09           & 4.35            \\
WavLM + NAW-SV \cite{lim2024improving}   & 102M+          & 2.96           & 3.29            \\
\textbf{ParaNoise-SV}           & \textbf{7.75M}  & 3.40  & 3.90   \\ \hline\hline
\end{tabular}
}
\vspace{-10pt}
\end{table}

\section{Conclusion}

ParaNoise-SV has been validated as an effective framework for noise-robust speaker verification by jointly optimizing noise extraction (NE), speech enhancement (SE), and speaker verification (SV). Its parallel network architecture enables dynamic noise suppression while preserving speaker-discriminative features. Experimental results across various SNRs and noise conditions show consistently lower EER, outperforming conventional SE-SV pipelines. The effectiveness of the dual U-Net architecture highlights the advantages of joint learning in improving speaker verification under real-world noisy conditions.

\pagebreak
\newpage

\section{Acknowledgements}

This work was conducted by Center for Applied Research in Artificial Intelligence(CARAI) grant funded by DAPA and ADD
(UD190031RD).

\bibliographystyle{IEEEtran}
\bibliography{mybib}

\end{document}